\documentclass[aps,prl,twocolumn]{revtex4-1}

\usepackage{graphicx}
\usepackage{bm}
\begin{document}

\title{A Novel Test of the Modified Newtonian Dynamics with Gas Rich Galaxies}

\author{Stacy S. McGaugh} 
\affiliation{Department of Astronomy, University of Maryland, College Park, MD 20742-2421}

\date{\today}

\begin{abstract}
The current cosmological paradigm, $\Lambda$CDM, requires that the mass-energy of the universe be dominated
by invisible components: dark matter and dark energy.  
An alternative to these dark components is that the law of gravity be modified on the relevant scales. 
A test of these ideas is provided by the Baryonic Tully-Fisher Relation (BTFR), an empirical relation between 
the observed mass of a galaxy and its rotation velocity.  Here I report a test using gas rich galaxies for which
both axes of the BTFR can be measured independently of the theories being tested and without the systematic
uncertainty in stellar mass that affects the same test with star dominated spirals.
The data fall precisely where predicted \textit{a priori} by the modified Newtonian dynamics (MOND).
The scatter in the BTFR is attributable entirely to observational uncertainty.  
This is consistent with the action of a single effective force law but poses a serious fine-tuning problem for $\Lambda$CDM.
\end{abstract}


\maketitle

The mass discrepancy problem in extragalactic systems is well established.   
When known dynamical laws are applied to these systems, 
the observed mass in stars and gas falls well short of explaining the observed motions.  A classic
example is that the rotation curves of disk galaxies tend to become roughly flat 
($V_f \sim$ constant) when they should be falling in a Keplerian ($V \propto r^{-1/2}$) fashion.  
The common interpretation for this phenomenon is dark matter.  
However, a logical alternative is that the dynamical laws that lead to the inference of dark matter 
need to be revised on the scales appropriate to galaxies.

One striking fact about extragalactic systems is that they are many orders of magnitude larger than the solar
system in which conventional dynamics is extraordinarily well tested.  One idea is thus to
modify gravity on some suitably large length scale such that the apparent need for dark matter would
be manifest in galaxies but not in the solar system.  Such size-dependent 
ideas fail and can generically be excluded \cite{MdB98a}.  
However, there are other ways in which galaxies differ from the solar system. 
For example, the centripetal acceleration required to keep a star in orbit in a galaxy
is very much lower than that experienced by the planets orbiting the sun:
$\sim 10^{-10}\;\textrm{m}\,\textrm{s}^{-2}$ vs.\ $6 \times 10^{-3}\;\textrm{m}\,\textrm{s}^{-2}$ for the Earth.

MOND \cite{MONDorig} posits a new constant with dimensions of acceleration, $a_0$,
which defines the boundary between conventional dynamics and a new domain of dynamics.
The conventional dynamics hold in the limit of high acceleration, $a \gg a_0$, 
and the modified regime occurs in the limit of low accelerations, $a \ll a_0$.
The value of $a_0$ must be determined observationally 
\cite{*[{The value $a_0 = 1.21 \pm 0.24 \times 10^{-10}\;\textrm{m}\,\textrm{s}^{-2}$ was obtained by }]
[{ from detailed fits to the rotation curves of star dominated galaxies.  Their procedure necessarily treats 
the mass-to-light ratio of each galaxy as a free parameter.  The use of gas rich galaxies here allows us to avoid 
this additional freedom.}] BBS}, but once specified is constant.  
In the modified regime, rotation curves  become asymptotically flat far from
a central mass \cite{MONDorig}.  This follows from the scale invariance symmetry of the equations of motion
under transformations  $(t, r) \rightarrow (\lambda t, \lambda r)$ \cite{scaleinvar}.
An absolute relation between the asymptotically flat rotation velocity $V_f$ and the total mass $M_b$
 \begin{equation}  a_0 G M_b = V_f^4 \label{eqn:mondTF} \end{equation}
follows uniquely on dimensional grounds.

Rotationally supported galaxies follow an empirical relation between mass and rotation velocity
known as the Baryonic Tully-Fisher Relation (BTFR) \cite{btforig}.  This empirical relation can,
in principle, provide a quantitative test of the prediction of MOND.  In order to do so, we require
independent, accurate measurements of both $M_b$ and $V_f$.  While the latter is readily obtained
from resolved rotation curves, mass determinations are more problematic.

The baryonic mass is the sum of both stars and gas: $M_b = M_{\star} + M_g$.  
A great deal is known about stars, but stellar mass estimates for galaxies 
are subject to a systematic uncertainty of $\sim 0.15$ dex because of
uncertainty in the stellar mass function and some details of stellar evolution \cite{MLerror}.
This level of systematic uncertainty precludes an unambiguous test of (\ref{eqn:mondTF}) 
with star dominated spirals \cite{M05}.

A clean test of the BTFR predicted by MOND follows if we can identify a class of galaxies where stars do not
dominate the baryonic mass budget.  Atomic gas typically dominates the mass of non-stellar material in disk galaxies.
Its mass follows directly from the distance to each galaxy, the measured 21 cm flux, 
and the physics of the spin flip transition of hydrogen.  
It does not suffer from the systematic uncertainty of stellar mass.  

\begin{figure*}
\includegraphics[]{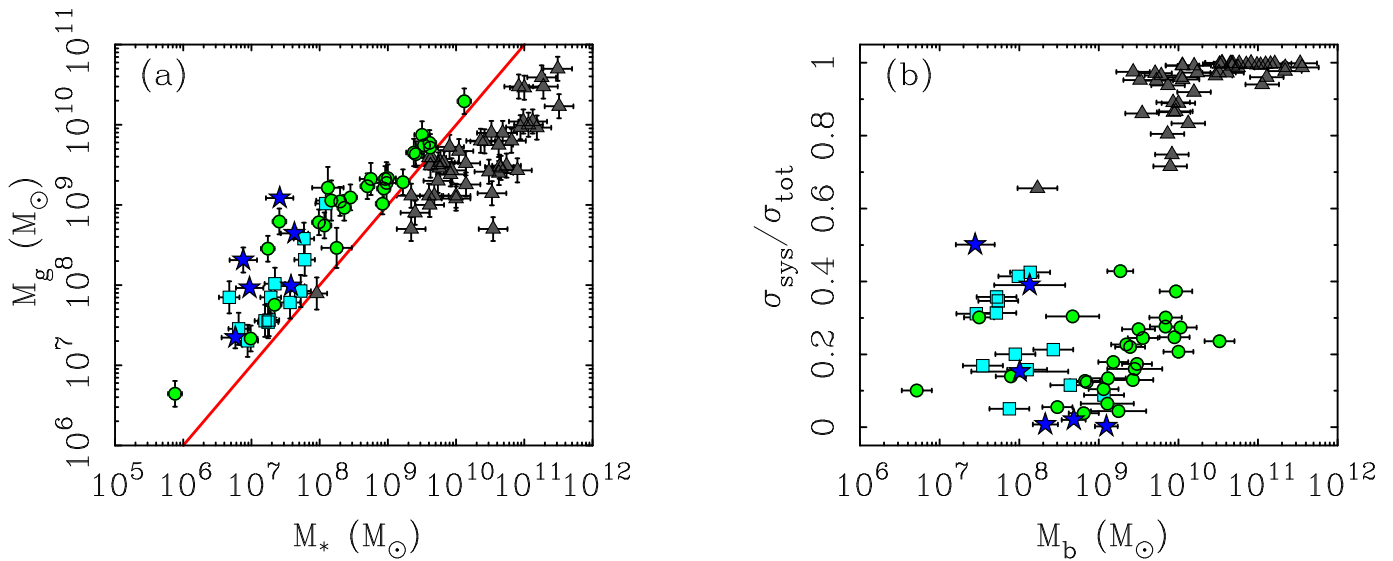}
\caption{(a) The mass of stars and gas in rotating galaxies.
Triangles represent star dominated spirals \cite{M05} with $M_{\star} > M_g$.
The data for gas rich galaxies with $M_{\star} < M_g$ come from several independent sources 
denoted by circles \cite{stark}, squares \cite{begum}, and stars \cite{trach}. 
(b) The fraction of the error budget $\sigma_{sys}/\sigma_{tot}$ contributed by the
systematic uncertainty in stellar mass ($\sigma_{sys} \approx 0.15$ dex \cite{MLerror})
as a function of the total baryonic mass $M_b = M_{\star} + M_g$.  Our knowledge of the masses of
star dominated galaxies is limited by this systematic uncertainty, but it has little effect on  
gas dominated galaxies.
\label{MstMg}}
\end{figure*}

Late type, low surface brightness disk galaxies frequently have gas masses in excess of their stellar 
masses \cite{gasfrac}.  When $M_g > M_{\star}$, the systematic uncertainty
in stellar mass is reduced to a minor contributor to the error budget (Fig.~\ref{MstMg}).
Thanks to recent work \cite{stark,begum,trach}, it is now possible to assemble a large
sample (47) of galaxies with $V_f$ measured from resolved rotation curves 
that satisfy the gas domination criterion $M_g > M_{\star}$.  
This property enables a novel test of MOND with no free parameters.
Both $M_b$ and $V_f$ are directly measured and, for the first time, are not dominated by systematic uncertainties.
Moreover, these galaxies are unambiguously in the deep modified regime
where (\ref{eqn:mondTF}) holds, with 
$V_f^2/r_{max} \approx  a_0/10$. 
Here the distinction between MOND and $\Lambda$CDM is most pronounced.

\begin{figure}
\includegraphics[]{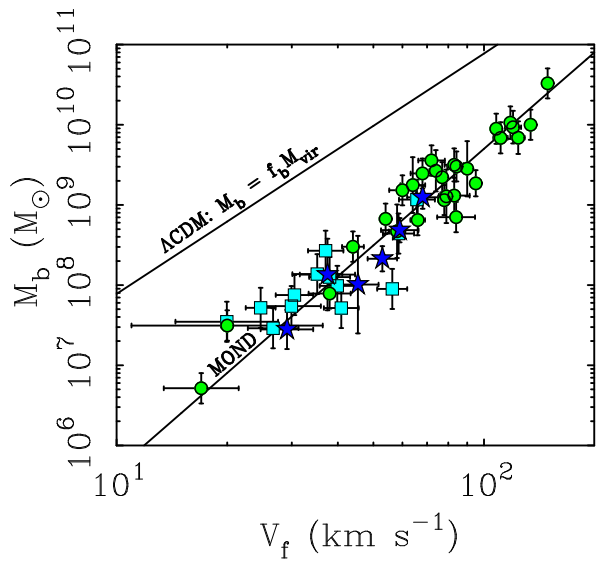}
\caption{The BTFR for gas dominated galaxies.
The sum of detected baryonic mass, stars and gas, is plotted against the flat rotation 
velocity $V_f$ (symbols as per Fig.~\ref{MstMg}).  Both mass and velocity are measured
independently of either MOND or $\Lambda$CDM.  The data are well removed from
the expectation of the standard cosmology (upper line), but
follow the prediction of MOND (lower line) with no fitting whatsoever.  
\label{btfmond}}
\end{figure}

Fig.~\ref{btfmond} shows the gas rich galaxy data together with the predictions of MOND and $\Lambda$CDM.
The data fall precisely where MOND predicts.  This happens with no fitting whatsoever --- there are \textit{zero} 
free parameters in Fig.~\ref{btfmond}.  Computing $\chi^2$ with the slope fixed to 4 and the normalization fixed
at the previously determined value of $a_0$ \cite{BBS}
gives $\chi^2 = 44.3$ for 46 degrees of freedom for a reduced $\chi_{\nu}^2 = 0.96$.
If we treat $a_0$ as a fit parameter and minimize $\chi^2$ with the slope fixed to 4, we find
$a_0 = 1.24 \pm 0.14 \times 10^{-10}\;\textrm{m}\,\textrm{s}^{-2}$.
This is indistinguishable from the previous value, and 
$\chi^2$ actually increases because we have
added an unneeded degree of freedom: $\chi_{\nu}^2 = 0.99$.
If we further treat the slope as an additional free parameter, we find $3.8 \pm 0.2$.
This does not differ significantly from the MOND prediction of 4, nor does it
improve the fit: $\chi_{\nu}^2 = 0.98$.  The data therefore provide
no reason to suspect a BTFR that differs in any way from that predicted by MOND.

The specific BTFR that the data follow is unique to MOND.
Indeed, to the best of my knowledge, MOND is the only theory to make a strong \textit{a priori} prediction
for the BTFR.  The dark matter paradigm makes no comparably iron-clad prediction.  

The expectation in $\Lambda$CDM is that total mass (both dark and baryonic) 
scales with rotation velocity as $M_{\Delta} = (\Delta/2)^{-1/2} (G H_0)^{-1} \; V_{\Delta}^3$.
These quantities are defined at a radius where the enclosed density
exceeds the cosmic critical density by a factor $\Delta$.  The virial radius occurs at
$\Delta \approx 100$ \cite{*[{}] [{. Numerically, the proportionality constant is
$4.6 \times 10^5\;\mathrm{M}_{\odot}\,\mathrm{km}^{-3}\,\mathrm{s}^3$ for 
$H_0 = 72\;\textrm{km}\,\textrm{s}^{-1}\,\textrm{Mpc}^{-1}$.}] MoMao}. 
This notional radius is well beyond the reach of observations.  
To plot the $\Lambda$CDM line in Fig.~\ref{btfmond}, we assume
$V_f = V_{vir}$ and $M_b = f_b M_{vir}$ where $f_b = 0.17$ is the cosmic baryon fraction \cite{WMAP5}.
This nominal expectation has the wrong slope and the wrong normalization.

In order to reconcile $\Lambda$CDM with the data, we must invoke additional parameters.
The simplest assumption is that only a fraction $f_d$ of the baryons in a halo are detected:
$M_b = f_d f_b M_{vir}$.  Once we have granted ourselves this freedom, a galaxy could, in principle, 
have any $f_d < 1$ and reside anywhere below the $\Lambda$CDM line in Fig.~\ref{btfmond}.  
From this perspective, it is puzzling that galaxies reside only along the line predicted by MOND.

Reproducing the observed BTFR in $\Lambda$CDM requires a remarkable degree of fine-tuning.  
The detected baryon fraction must follow the formula 
$\log f_d = \log (V_f/100\;\mathrm{km}\,\mathrm{s}^{-1}) - 1.2$.  
Astrophysical feedback is often invoked to cause $f_d < 1$, but provides no satisfactory explanation
for why this particular tuning of $f_d$ arises.  

A further test is provided by the scatter in the observed relation.  
In $\Lambda$CDM, any scatter in $f_d$ translates directly into the BTFR: 
there should be at least some intrinsic scatter.  In MOND, the BTFR
is a consequence of the force law, and should have no intrinsic scatter.

\begin{figure}
\includegraphics[]{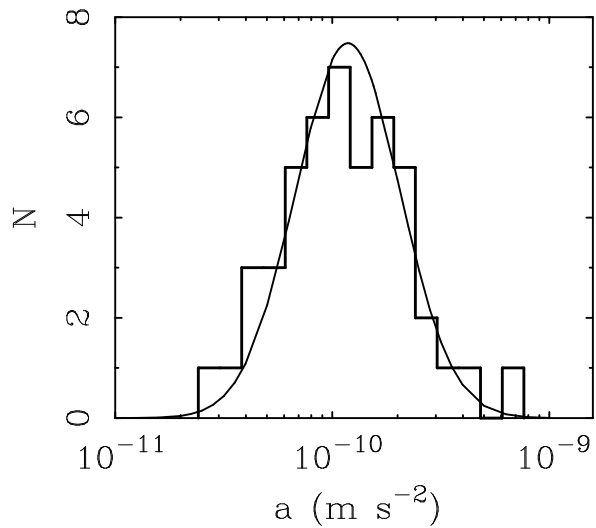}
\caption{Histogram of the measured values $\mathrm{a} = V_f^4/(GM_b)$.
The data are consistent with a normal distribution (smooth curve) that 
is centered on the previous determination of $a_0$ \cite{BBS}
with a width specified by the mean uncertainty $\sigma = 0.24$ dex.
This consistency implies a universal acceleration scale with negligible
intrinsic scatter.  This is expected in MOND, but poses a fine-tuning problem for $\Lambda$CDM.
\label{azero}}
\end{figure}

Fig.~\ref{azero} shows a histogram of the ratio $\mathrm{a}  = V_f^4/(G M_b)$ formed from the data.  
This represents the scatter around the MOND line in Fig.~\ref{btfmond}.  If the data are randomly distributed,
they should approximate a gaussian whose width is dictated by the size of the error bars.  
Such a gaussian is shown in Fig.~\ref{azero}.  It is not fit to the data; it is simply centered on the previously 
determined value of $a_0$ \cite{BBS} with a width corresponding to the uncertainty in
the data.  Only random errors are considered here; the residual systematic uncertainty in the 
stellar mass corresponds to a small shift in the total mass and should not introduce additional scatter. 

Observational uncertainty suffices to explain the scatter in the data.  
The data are consistent with a BTFR of zero intrinsic width.  
This is natural if the BTFR is imposed by the force law, as in MOND.  
It is not expected in $\Lambda$CDM where there should be many sources of scatter.

From the perspective of cosmology, it is disturbing that MOND works at all.
If $\Lambda$CDM is the correct paradigm, this should not happen \cite{sandersNOCDM}.
Yet when pressed into a new regime where the predictions of the two theories
are distinct, MOND outperforms $\Lambda$CDM.

This is not the first time that strong predictions of MOND have been realized.  
For example, MOND predicted in advance that galaxies of both high and low surface brightness would
fall on the same BTFR \cite{zwaanTF,sprayTF}, contrary to the natural 
expectation of purely Newtonian gravity \cite{MdB98b,myPRL}.  It is well established that
MOND provides good fits to the detailed shapes of rotation curves with only the stellar mass-to-light ratio as a 
free parameter \cite{SMmond}.  The required mass-to-light ratios are in good agreement
with stellar population synthesis models \cite{Iran}.  A simple model motivated by 
MOND provided the only successful \textit{a priori} prediction of the first-to-second peak amplitude 
ratio of the acoustic peaks of the cosmic background radiation:  $A_{1:2} = 2.4$ 
predicted \cite{CMB99} vs.\ $2.34 \pm 0.09$ measured \cite{wmappeaks}.  
It is rare for a non-canonical theory to have so many predictive successes.

MOND also has its share of problems.
The same ansatz that correctly predicted the second acoustic peak amplitude
also predicts a lower third peak than is observed \cite{CMB04}.  This does not falsify MOND,
but it does imply that a generally covariant parent theory should provide an effective forcing term \cite{TVSforcing}.

The most serious observational problem facing MOND is the dynamics of rich clusters of galaxies.
These appear to weigh more than can be accounted for with the observed baryons
even with the modified dynamics \cite{sandersclusters,angus}.
This residual mass discrepancy is roughly a factor of two in mass.  On the one hand, this is very disturbing ---
a theory that seeks to eliminate the need for cosmic dark matter itself suffers a missing mass problem.
On the other hand, this is less severe than the missing baryon problem in $\Lambda$CDM,
where dwarf galaxies are missing 99\% of the baryons that should be associated with
their dark matter halos \cite{M10}.  So both theories suffer a missing baryon problem, albeit
of different amplitudes in systems of vastly different scale.

While some of the mass in clusters appears to be dark, even in MOND, 
there is nothing that requires this unseen mass to be in some new form of 
non-baryonic particle.  Indeed, big bang nucleosynthesis implies the existence of considerably
more baryons than have so far been detected \cite{fukugita}.
If only a fraction of these missing baryons reside in clusters it would
suffice to resolve the residual mass discrepancy suffered by MOND.

Perhaps the most prominent example of a cluster with a serious residual discrepancy in MOND
is the bullet cluster \cite{clowe}.  In this system, the gravitational lensing of background galaxies 
indicates that the mass is offset from the X-ray plasma.  This is the same residual
mass discrepancy that is seen in all rich clusters.  While the bullet cluster 
is frequently cited as evidence against MOND, it is also problematic for $\Lambda$CDM.
The sub-clusters that compose the bullet cluster collided at a remarkably high velocity 
($\sim 4700\;\mathrm{km}\,\mathrm{s}^{-1}$). 
This is exceedingly unlikely in $\Lambda$CDM, occurring with a probability of
only a few parts in a billion \cite{bulletspeed}.  In contrast, such high collision velocities are natural
to MOND \cite{angmcg}.  Taken at face value, the bullet cluster would seem to 
simultaneously support and falsify both theories with equal vigor.

Given the nature of astronomical data, some exceptions to any theory are to be expected.
What is surprising in the case of MOND is that it continues to enjoy predictive successes at all.
These motivate the search for a more complete gravitational theory that contains MOND in the appropriate 
limit \cite{TeVeS,brownsteinmoffat,bimond,mannheimobrien}.  

It is possible that non-baryonic cold dark matter does not exist.  
If it does, and $\Lambda$CDM is the correct solution, the challenge is to understand the 
empirical systematics encapsulated in the simple MOND formula.
These are not native to the current cosmological paradigm \cite{sandersNOCDM}
but must be explained by any successful theory.

Another possibility is that dark matter particles have properties that impose MOND-like 
phenomenology \cite{blanchet,hybridDM,darkfluid}.  In this case, it is desirable to have dark matter
that behaves like standard cold dark matter on large scales, but which interacts with normal matter
so as to impose the MOND phenomenology in galaxies.  This suggests some strong new form of
interaction between dark matter and baryons. 

If MOND is essentially correct in that it is pointing towards an extension of gravitational theory, then
the experiments seeking to detect dark matter will find null results.  This would also be the case if 
dark matter has a different nature than currently presumed.  If dark matter is detected, 
then the MOND formula is still useful as a phenomenological constraint on the effective
force law in spiral galaxies.  In any case, the predictive power of the simple formula proposed 
by Milgrom \cite{MONDorig} is telling us something profound about Nature.

\begin{acknowledgements}
I thank Benoit Famaey for organizing the conference where these issues came into focus.
I also thank Moti Milgrom and Bob Sanders for sharing their perspectives, and the referees
for their constructive comments.  The work of SSM is supported in part by NSF grant AST0908370.
\end{acknowledgements}

\bibliography{PRL_ms}

\begin{thebibliography}{38}%
\makeatletter
\providecommand \@ifxundefined [1]{%
 \@ifx{#1\undefined}
}%
\providecommand \@ifnum [1]{%
 \ifnum #1\expandafter \@firstoftwo
 \else \expandafter \@secondoftwo
 \fi
}%
\providecommand \@ifx [1]{%
 \ifx #1\expandafter \@firstoftwo
 \else \expandafter \@secondoftwo
 \fi
}%
\providecommand \natexlab [1]{#1}%
\providecommand \enquote  [1]{``#1''}%
\providecommand \bibnamefont  [1]{#1}%
\providecommand \bibfnamefont [1]{#1}%
\providecommand \citenamefont [1]{#1}%
\providecommand \href@noop [0]{\@secondoftwo}%
\providecommand \href [0]{\begingroup \@sanitize@url \@href}%
\providecommand \@href[1]{\@@startlink{#1}\@@href}%
\providecommand \@@href[1]{\endgroup#1\@@endlink}%
\providecommand \@sanitize@url [0]{\catcode `\\12\catcode `\$12\catcode
  `\&12\catcode `\#12\catcode `\^12\catcode `\_12\catcode `\%12\relax}%
\providecommand \@@startlink[1]{}%
\providecommand \@@endlink[0]{}%
\providecommand \url  [0]{\begingroup\@sanitize@url \@url }%
\providecommand \@url [1]{\endgroup\@href {#1}{\urlprefix }}%
\providecommand \urlprefix  [0]{URL }%
\providecommand \Eprint [0]{\href }%
\providecommand \doibase [0]{http://dx.doi.org/}%
\providecommand \selectlanguage [0]{\@gobble}%
\providecommand \bibinfo  [0]{\@secondoftwo}%
\providecommand \bibfield  [0]{\@secondoftwo}%
\providecommand \translation [1]{[#1]}%
\providecommand \BibitemOpen [0]{}%
\providecommand \bibitemStop [0]{}%
\providecommand \bibitemNoStop [0]{.\EOS\space}%
\providecommand \EOS [0]{\spacefactor3000\relax}%
\providecommand \BibitemShut  [1]{\csname bibitem#1\endcsname}%
\let\auto@bib@innerbib\@empty
\bibitem [{\citenamefont {{McGaugh}}\ and\ \citenamefont {{de
  Blok}}(1998{\natexlab{a}})}]{MdB98a}%
  \BibitemOpen
  \bibfield  {author} {\bibinfo {author} {\bibfnamefont {S.~S.}\ \bibnamefont
  {{McGaugh}}}\ and\ \bibinfo {author} {\bibfnamefont {W.~J.~G.}\ \bibnamefont
  {{de Blok}}},\ }\href@noop {} {\bibfield  {journal} {\bibinfo  {journal}
  {Astrophys. J.}\ }\textbf {\bibinfo {volume} {499}},\ \bibinfo {pages} {41}
  (\bibinfo {year} {1998}{\natexlab{a}})}\BibitemShut {NoStop}%
\bibitem [{\citenamefont {{Milgrom}}(1983)}]{MONDorig}%
  \BibitemOpen
  \bibfield  {author} {\bibinfo {author} {\bibfnamefont {M.}~\bibnamefont
  {{Milgrom}}},\ }\href {\doibase 10.1086/161130} {\bibfield  {journal}
  {\bibinfo  {journal} {Astrophys. J.}\ }\textbf {\bibinfo {volume} {270}},\
  \bibinfo {pages} {365} (\bibinfo {year} {1983})}\BibitemShut {NoStop}%
\bibitem [{\citenamefont {{Begeman}}\ \emph {et~al.}(1991)\citenamefont
  {{Begeman}}, \citenamefont {{Broeils}},\ and\ \citenamefont
  {{Sanders}}}]{BBS}%
  \BibitemOpen
  \bibfield  {author} {\bibinfo {author} {\bibfnamefont {K.~G.}\ \bibnamefont
  {{Begeman}}}, \bibinfo {author} {\bibfnamefont {A.~H.}\ \bibnamefont
  {{Broeils}}}, \ and\ \bibinfo {author} {\bibfnamefont {R.~H.}\ \bibnamefont
  {{Sanders}}},\ }\href@noop {} {\bibfield  {journal} {\bibinfo  {journal}
  {Mon. Not. R. Astron. Soc.}\ }\textbf {\bibinfo {volume} {249}},\ \bibinfo
  {pages} {523} (\bibinfo {year} {1991})}\BibitemShut {NoStop}%
\bibitem [{\citenamefont {{Milgrom}}(2009{\natexlab{a}})}]{scaleinvar}%
  \BibitemOpen
  \bibfield  {author} {\bibinfo {author} {\bibfnamefont {M.}~\bibnamefont
  {{Milgrom}}},\ }\href {\doibase 10.1088/0004-637X/698/2/1630} {\bibfield
  {journal} {\bibinfo  {journal} {Astrophys. J.}\ }\textbf {\bibinfo {volume}
  {698}},\ \bibinfo {pages} {1630} (\bibinfo {year}
  {2009}{\natexlab{a}})}\BibitemShut {NoStop}%
\bibitem [{\citenamefont {{McGaugh}}\ \emph {et~al.}(2000)\citenamefont
  {{McGaugh}}, \citenamefont {{Schombert}}, \citenamefont {{Bothun}},\ and\
  \citenamefont {{de Blok}}}]{btforig}%
  \BibitemOpen
  \bibfield  {author} {\bibinfo {author} {\bibfnamefont {S.~S.}\ \bibnamefont
  {{McGaugh}}}, \bibinfo {author} {\bibfnamefont {J.~M.}\ \bibnamefont
  {{Schombert}}}, \bibinfo {author} {\bibfnamefont {G.~D.}\ \bibnamefont
  {{Bothun}}}, \ and\ \bibinfo {author} {\bibfnamefont {W.~J.~G.}\ \bibnamefont
  {{de Blok}}},\ }\href {\doibase 10.1086/312628} {\bibfield  {journal}
  {\bibinfo  {journal} {Astrophys. J.}\ }\textbf {\bibinfo {volume} {533}},\
  \bibinfo {pages} {L99} (\bibinfo {year} {2000})}\BibitemShut {NoStop}%
\bibitem [{\citenamefont {{Conroy}}\ \emph {et~al.}(2009)\citenamefont
  {{Conroy}}, \citenamefont {{Gunn}},\ and\ \citenamefont {{White}}}]{MLerror}%
  \BibitemOpen
  \bibfield  {author} {\bibinfo {author} {\bibfnamefont {C.}~\bibnamefont
  {{Conroy}}}, \bibinfo {author} {\bibfnamefont {J.~E.}\ \bibnamefont
  {{Gunn}}}, \ and\ \bibinfo {author} {\bibfnamefont {M.}~\bibnamefont
  {{White}}},\ }\href {\doibase 10.1088/0004-637X/699/1/486} {\bibfield
  {journal} {\bibinfo  {journal} {Astrophys. J.}\ }\textbf {\bibinfo {volume}
  {699}},\ \bibinfo {pages} {486} (\bibinfo {year} {2009})}\BibitemShut
  {NoStop}%
\bibitem [{\citenamefont {{McGaugh}}(2005{\natexlab{a}})}]{M05}%
  \BibitemOpen
  \bibfield  {author} {\bibinfo {author} {\bibfnamefont {S.~S.}\ \bibnamefont
  {{McGaugh}}},\ }\href {\doibase 10.1086/432968} {\bibfield  {journal}
  {\bibinfo  {journal} {Astrophys. J.}\ }\textbf {\bibinfo {volume} {632}},\
  \bibinfo {pages} {859} (\bibinfo {year} {2005}{\natexlab{a}})}\BibitemShut
  {NoStop}%
\bibitem [{\citenamefont {{Schombert}}\ \emph {et~al.}(2001)\citenamefont
  {{Schombert}}, \citenamefont {{McGaugh}},\ and\ \citenamefont
  {{Eder}}}]{gasfrac}%
  \BibitemOpen
  \bibfield  {author} {\bibinfo {author} {\bibfnamefont {J.~M.}\ \bibnamefont
  {{Schombert}}}, \bibinfo {author} {\bibfnamefont {S.~S.}\ \bibnamefont
  {{McGaugh}}}, \ and\ \bibinfo {author} {\bibfnamefont {J.~A.}\ \bibnamefont
  {{Eder}}},\ }\href {\doibase 10.1086/320398} {\bibfield  {journal} {\bibinfo
  {journal} {Astron. J.}\ }\textbf {\bibinfo {volume} {121}},\ \bibinfo {pages}
  {2420} (\bibinfo {year} {2001})}\BibitemShut {NoStop}%
\bibitem [{\citenamefont {{Stark}}\ \emph {et~al.}(2009)\citenamefont
  {{Stark}}, \citenamefont {{McGaugh}},\ and\ \citenamefont
  {{Swaters}}}]{stark}%
  \BibitemOpen
  \bibfield  {author} {\bibinfo {author} {\bibfnamefont {D.~V.}\ \bibnamefont
  {{Stark}}}, \bibinfo {author} {\bibfnamefont {S.~S.}\ \bibnamefont
  {{McGaugh}}}, \ and\ \bibinfo {author} {\bibfnamefont {R.~A.}\ \bibnamefont
  {{Swaters}}},\ }\href {\doibase 10.1088/0004-6256/138/2/392} {\bibfield
  {journal} {\bibinfo  {journal} {Astron. J.}\ }\textbf {\bibinfo {volume}
  {138}},\ \bibinfo {pages} {392} (\bibinfo {year} {2009})}\BibitemShut
  {NoStop}%
\bibitem [{\citenamefont {{Begum}}\ \emph {et~al.}(2008)\citenamefont
  {{Begum}}, \citenamefont {{Chengalur}}, \citenamefont {{Karachentsev}},\ and\
  \citenamefont {{Sharina}}}]{begum}%
  \BibitemOpen
  \bibfield  {author} {\bibinfo {author} {\bibfnamefont {A.}~\bibnamefont
  {{Begum}}}, \bibinfo {author} {\bibfnamefont {J.~N.}\ \bibnamefont
  {{Chengalur}}}, \bibinfo {author} {\bibfnamefont {I.~D.}\ \bibnamefont
  {{Karachentsev}}}, \ and\ \bibinfo {author} {\bibfnamefont {M.~E.}\
  \bibnamefont {{Sharina}}},\ }\href {\doibase
  10.1111/j.1365-2966.2008.13010.x} {\bibfield  {journal} {\bibinfo  {journal}
  {Mon. Not. R. Astron. Soc.}\ }\textbf {\bibinfo {volume} {386}},\ \bibinfo
  {pages} {138} (\bibinfo {year} {2008})}\BibitemShut {NoStop}%
\bibitem [{\citenamefont {{Trachternach}}\ \emph {et~al.}(2009)\citenamefont
  {{Trachternach}}, \citenamefont {{de Blok}}, \citenamefont {{McGaugh}},
  \citenamefont {{van der Hulst}},\ and\ \citenamefont {{Dettmar}}}]{trach}%
  \BibitemOpen
  \bibfield  {author} {\bibinfo {author} {\bibfnamefont {C.}~\bibnamefont
  {{Trachternach}}}, \bibinfo {author} {\bibfnamefont {W.~J.~G.}\ \bibnamefont
  {{de Blok}}}, \bibinfo {author} {\bibfnamefont {S.~S.}\ \bibnamefont
  {{McGaugh}}}, \bibinfo {author} {\bibfnamefont {J.~M.}\ \bibnamefont {{van
  der Hulst}}}, \ and\ \bibinfo {author} {\bibfnamefont {R.}~\bibnamefont
  {{Dettmar}}},\ }\href {\doibase 10.1051/0004-6361/200811136} {\bibfield
  {journal} {\bibinfo  {journal} {Astron. Astrophys.}\ }\textbf {\bibinfo
  {volume} {505}},\ \bibinfo {pages} {577} (\bibinfo {year}
  {2009})}\BibitemShut {NoStop}%
\bibitem [{\citenamefont {{Mo}}\ and\ \citenamefont {{Mao}}(2004)}]{MoMao}%
  \BibitemOpen
  \bibfield  {author} {\bibinfo {author} {\bibfnamefont {H.~J.}\ \bibnamefont
  {{Mo}}}\ and\ \bibinfo {author} {\bibfnamefont {S.}~\bibnamefont {{Mao}}},\
  }\href@noop {} {\bibfield  {journal} {\bibinfo  {journal} {Mon. Not. R.
  Astron. Soc.}\ }\textbf {\bibinfo {volume} {353}},\ \bibinfo {pages} {829}
  (\bibinfo {year} {2004})}\BibitemShut {NoStop}%
\bibitem [{\citenamefont {{Komatsu \textit{et al.}}}(2009)}]{WMAP5}%
  \BibitemOpen
  \bibfield  {author} {\bibinfo {author} {\bibfnamefont {E.}~\bibnamefont
  {{Komatsu \textit{et al.}}}},\ }\href {\doibase 10.1088/0067-0049/180/2/330}
  {\bibfield  {journal} {\bibinfo  {journal} {Astrophys. J. Suppl.}\ }\textbf
  {\bibinfo {volume} {180}},\ \bibinfo {pages} {330} (\bibinfo {year}
  {2009})}\BibitemShut {NoStop}%
\bibitem [{\citenamefont {{Sanders}}(2009)}]{sandersNOCDM}%
  \BibitemOpen
  \bibfield  {author} {\bibinfo {author} {\bibfnamefont {R.~H.}\ \bibnamefont
  {{Sanders}}},\ }\href {\doibase 10.1155/2009/752439} {\bibfield  {journal}
  {\bibinfo  {journal} {Advances in Astronomy}\ }\textbf {\bibinfo {volume}
  {2009}},\ \bibinfo {pages} {752439} (\bibinfo {year} {2009})}\BibitemShut
  {NoStop}%
\bibitem [{\citenamefont {{Zwaan}}\ \emph {et~al.}(1995)\citenamefont
  {{Zwaan}}, \citenamefont {{van der Hulst}}, \citenamefont {{de Blok}},\ and\
  \citenamefont {{McGaugh}}}]{zwaanTF}%
  \BibitemOpen
  \bibfield  {author} {\bibinfo {author} {\bibfnamefont {M.~A.}\ \bibnamefont
  {{Zwaan}}}, \bibinfo {author} {\bibfnamefont {J.~M.}\ \bibnamefont {{van der
  Hulst}}}, \bibinfo {author} {\bibfnamefont {W.~J.~G.}\ \bibnamefont {{de
  Blok}}}, \ and\ \bibinfo {author} {\bibfnamefont {S.~S.}\ \bibnamefont
  {{McGaugh}}},\ }\href@noop {} {\bibfield  {journal} {\bibinfo  {journal}
  {Mon. Not. R. Astron. Soc.}\ }\textbf {\bibinfo {volume} {273}},\ \bibinfo
  {pages} {L35} (\bibinfo {year} {1995})}\BibitemShut {NoStop}%
\bibitem [{\citenamefont {{Sprayberry}}\ \emph {et~al.}(1995)\citenamefont
  {{Sprayberry}}, \citenamefont {{Bernstein}}, \citenamefont {{Impey}},\ and\
  \citenamefont {{Bothun}}}]{sprayTF}%
  \BibitemOpen
  \bibfield  {author} {\bibinfo {author} {\bibfnamefont {D.}~\bibnamefont
  {{Sprayberry}}}, \bibinfo {author} {\bibfnamefont {G.~M.}\ \bibnamefont
  {{Bernstein}}}, \bibinfo {author} {\bibfnamefont {C.~D.}\ \bibnamefont
  {{Impey}}}, \ and\ \bibinfo {author} {\bibfnamefont {G.~D.}\ \bibnamefont
  {{Bothun}}},\ }\href {\doibase 10.1086/175055} {\bibfield  {journal}
  {\bibinfo  {journal} {Astrophys. J.}\ }\textbf {\bibinfo {volume} {438}},\
  \bibinfo {pages} {72} (\bibinfo {year} {1995})}\BibitemShut {NoStop}%
\bibitem [{\citenamefont {{McGaugh}}\ and\ \citenamefont {{de
  Blok}}(1998{\natexlab{b}})}]{MdB98b}%
  \BibitemOpen
  \bibfield  {author} {\bibinfo {author} {\bibfnamefont {S.~S.}\ \bibnamefont
  {{McGaugh}}}\ and\ \bibinfo {author} {\bibfnamefont {W.~J.~G.}\ \bibnamefont
  {{de Blok}}},\ }\href@noop {} {\bibfield  {journal} {\bibinfo  {journal}
  {Astrophys. J.}\ }\textbf {\bibinfo {volume} {499}},\ \bibinfo {pages} {66}
  (\bibinfo {year} {1998}{\natexlab{b}})}\BibitemShut {NoStop}%
\bibitem [{\citenamefont {{McGaugh}}(2005{\natexlab{b}})}]{myPRL}%
  \BibitemOpen
  \bibfield  {author} {\bibinfo {author} {\bibfnamefont {S.~S.}\ \bibnamefont
  {{McGaugh}}},\ }\href {\doibase 10.1103/PhysRevLett.95.171302} {\bibfield
  {journal} {\bibinfo  {journal} {Phys. Rev. Lett.}\ }\textbf {\bibinfo
  {volume} {95}},\ \bibinfo {pages} {171302} (\bibinfo {year}
  {2005}{\natexlab{b}})}\BibitemShut {NoStop}%
\bibitem [{\citenamefont {{Sanders}}\ and\ \citenamefont
  {{McGaugh}}(2002)}]{SMmond}%
  \BibitemOpen
  \bibfield  {author} {\bibinfo {author} {\bibfnamefont {R.~H.}\ \bibnamefont
  {{Sanders}}}\ and\ \bibinfo {author} {\bibfnamefont {S.~S.}\ \bibnamefont
  {{McGaugh}}},\ }\href {\doibase 10.1146/annurev.astro.40.060401.093923}
  {\bibfield  {journal} {\bibinfo  {journal} {Ann. Rev. Astron. Astrophys.}\
  }\textbf {\bibinfo {volume} {40}},\ \bibinfo {pages} {263} (\bibinfo {year}
  {2002})}\BibitemShut {NoStop}%
\bibitem [{\citenamefont {{Hasani Zonoozi}}\ and\ \citenamefont
  {{Haghi}}(2010)}]{Iran}%
  \BibitemOpen
  \bibfield  {author} {\bibinfo {author} {\bibfnamefont {A.}~\bibnamefont
  {{Hasani Zonoozi}}}\ and\ \bibinfo {author} {\bibfnamefont {H.}~\bibnamefont
  {{Haghi}}},\ }\href {\doibase 10.1051/0004-6361/201014933} {\bibfield
  {journal} {\bibinfo  {journal} {Astron. Astrophys.}\ }\textbf {\bibinfo
  {volume} {524}},\ \bibinfo {pages} {A53} (\bibinfo {year}
  {2010})}\BibitemShut {NoStop}%
\bibitem [{\citenamefont {{McGaugh}}(1999)}]{CMB99}%
  \BibitemOpen
  \bibfield  {author} {\bibinfo {author} {\bibfnamefont {S.~S.}\ \bibnamefont
  {{McGaugh}}},\ }\href {\doibase 10.1086/312274} {\bibfield  {journal}
  {\bibinfo  {journal} {Astrophys. J.}\ }\textbf {\bibinfo {volume} {523}},\
  \bibinfo {pages} {L99} (\bibinfo {year} {1999})}\BibitemShut {NoStop}%
\bibitem [{\citenamefont {{Page \textit{et al.}}}(2003)}]{wmappeaks}%
  \BibitemOpen
  \bibfield  {author} {\bibinfo {author} {\bibfnamefont {L.}~\bibnamefont
  {{Page \textit{et al.}}}},\ }\href {\doibase 10.1086/377223} {\bibfield
  {journal} {\bibinfo  {journal} {Astrophys. J. Suppl.}\ }\textbf {\bibinfo
  {volume} {148}},\ \bibinfo {pages} {39} (\bibinfo {year} {2003})}\BibitemShut
  {NoStop}%
\bibitem [{\citenamefont {{McGaugh}}(2004)}]{CMB04}%
  \BibitemOpen
  \bibfield  {author} {\bibinfo {author} {\bibfnamefont {S.~S.}\ \bibnamefont
  {{McGaugh}}},\ }\href {\doibase 10.1086/421895} {\bibfield  {journal}
  {\bibinfo  {journal} {Astrophys. J.}\ }\textbf {\bibinfo {volume} {611}},\
  \bibinfo {pages} {26} (\bibinfo {year} {2004})}\BibitemShut {NoStop}%
\bibitem [{\citenamefont {{Skordis}}\ \emph {et~al.}(2006)\citenamefont
  {{Skordis}}, \citenamefont {{Mota}}, \citenamefont {{Ferreira}},\ and\
  \citenamefont {{B{\oe}hm}}}]{TVSforcing}%
  \BibitemOpen
  \bibfield  {author} {\bibinfo {author} {\bibfnamefont {C.}~\bibnamefont
  {{Skordis}}}, \bibinfo {author} {\bibfnamefont {D.~F.}\ \bibnamefont
  {{Mota}}}, \bibinfo {author} {\bibfnamefont {P.~G.}\ \bibnamefont
  {{Ferreira}}}, \ and\ \bibinfo {author} {\bibfnamefont {C.}~\bibnamefont
  {{B{\oe}hm}}},\ }\href {\doibase 10.1103/PhysRevLett.96.011301} {\bibfield
  {journal} {\bibinfo  {journal} {Physical Review Letters}\ }\textbf {\bibinfo
  {volume} {96}},\ \bibinfo {pages} {011301} (\bibinfo {year}
  {2006})}\BibitemShut {NoStop}%
\bibitem [{\citenamefont {{Sanders}}(2003)}]{sandersclusters}%
  \BibitemOpen
  \bibfield  {author} {\bibinfo {author} {\bibfnamefont {R.~H.}\ \bibnamefont
  {{Sanders}}},\ }\href {\doibase 10.1046/j.1365-8711.2003.06596.x} {\bibfield
  {journal} {\bibinfo  {journal} {Mon. Not R. Astron. Soc.}\ }\textbf {\bibinfo
  {volume} {342}},\ \bibinfo {pages} {901} (\bibinfo {year}
  {2003})}\BibitemShut {NoStop}%
\bibitem [{\citenamefont {{Angus}}\ \emph {et~al.}(2008)\citenamefont
  {{Angus}}, \citenamefont {{Famaey}},\ and\ \citenamefont {{Buote}}}]{angus}%
  \BibitemOpen
  \bibfield  {author} {\bibinfo {author} {\bibfnamefont {G.~W.}\ \bibnamefont
  {{Angus}}}, \bibinfo {author} {\bibfnamefont {B.}~\bibnamefont {{Famaey}}}, \
  and\ \bibinfo {author} {\bibfnamefont {D.~A.}\ \bibnamefont {{Buote}}},\
  }\href {\doibase 10.1111/j.1365-2966.2008.13353.x} {\bibfield  {journal}
  {\bibinfo  {journal} {Mon. Not. R. Astron. Soc.}\ }\textbf {\bibinfo {volume}
  {387}},\ \bibinfo {pages} {1470} (\bibinfo {year} {2008})}\BibitemShut
  {NoStop}%
\bibitem [{\citenamefont {{McGaugh}}\ \emph {et~al.}(2010)\citenamefont
  {{McGaugh}}, \citenamefont {{Schombert}}, \citenamefont {{de Blok}},\ and\
  \citenamefont {{Zagursky}}}]{M10}%
  \BibitemOpen
  \bibfield  {author} {\bibinfo {author} {\bibfnamefont {S.~S.}\ \bibnamefont
  {{McGaugh}}}, \bibinfo {author} {\bibfnamefont {J.~M.}\ \bibnamefont
  {{Schombert}}}, \bibinfo {author} {\bibfnamefont {W.~J.~G.}\ \bibnamefont
  {{de Blok}}}, \ and\ \bibinfo {author} {\bibfnamefont {M.~J.}\ \bibnamefont
  {{Zagursky}}},\ }\href {\doibase 10.1088/2041-8205/708/1/L14} {\bibfield
  {journal} {\bibinfo  {journal} {Astrophys. J.}\ }\textbf {\bibinfo {volume}
  {708}},\ \bibinfo {pages} {L14} (\bibinfo {year} {2010})}\BibitemShut
  {NoStop}%
\bibitem [{\citenamefont {{Fukugita}}\ \emph {et~al.}(1998)\citenamefont
  {{Fukugita}}, \citenamefont {{Hogan}},\ and\ \citenamefont
  {{Peebles}}}]{fukugita}%
  \BibitemOpen
  \bibfield  {author} {\bibinfo {author} {\bibfnamefont {M.}~\bibnamefont
  {{Fukugita}}}, \bibinfo {author} {\bibfnamefont {C.~J.}\ \bibnamefont
  {{Hogan}}}, \ and\ \bibinfo {author} {\bibfnamefont {P.~J.~E.}\ \bibnamefont
  {{Peebles}}},\ }\href@noop {} {\bibfield  {journal} {\bibinfo  {journal}
  {Astrophys. J.}\ }\textbf {\bibinfo {volume} {503}},\ \bibinfo {pages} {518}
  (\bibinfo {year} {1998})}\BibitemShut {NoStop}%
\bibitem [{\citenamefont {{Clowe}}\ \emph {et~al.}(2004)\citenamefont
  {{Clowe}}, \citenamefont {{Gonzalez}},\ and\ \citenamefont
  {{Markevitch}}}]{clowe}%
  \BibitemOpen
  \bibfield  {author} {\bibinfo {author} {\bibfnamefont {D.}~\bibnamefont
  {{Clowe}}}, \bibinfo {author} {\bibfnamefont {A.}~\bibnamefont {{Gonzalez}}},
  \ and\ \bibinfo {author} {\bibfnamefont {M.}~\bibnamefont {{Markevitch}}},\
  }\href {\doibase 10.1086/381970} {\bibfield  {journal} {\bibinfo  {journal}
  {Astrophys. J.}\ }\textbf {\bibinfo {volume} {604}},\ \bibinfo {pages} {596}
  (\bibinfo {year} {2004})}\BibitemShut {NoStop}%
\bibitem [{\citenamefont {{Lee}}\ and\ \citenamefont
  {{Komatsu}}(2010)}]{bulletspeed}%
  \BibitemOpen
  \bibfield  {author} {\bibinfo {author} {\bibfnamefont {J.}~\bibnamefont
  {{Lee}}}\ and\ \bibinfo {author} {\bibfnamefont {E.}~\bibnamefont
  {{Komatsu}}},\ }\href {\doibase 10.1088/0004-637X/718/1/60} {\bibfield
  {journal} {\bibinfo  {journal} {Astrophys. J.}\ }\textbf {\bibinfo {volume}
  {718}},\ \bibinfo {pages} {60} (\bibinfo {year} {2010})}\BibitemShut
  {NoStop}%
\bibitem [{\citenamefont {{Angus}}\ and\ \citenamefont
  {{McGaugh}}(2008)}]{angmcg}%
  \BibitemOpen
  \bibfield  {author} {\bibinfo {author} {\bibfnamefont {G.~W.}\ \bibnamefont
  {{Angus}}}\ and\ \bibinfo {author} {\bibfnamefont {S.~S.}\ \bibnamefont
  {{McGaugh}}},\ }\href {\doibase 10.1111/j.1365-2966.2007.12403.x} {\bibfield
  {journal} {\bibinfo  {journal} {Mon. Not. R. Astron. Soc.}\ }\textbf
  {\bibinfo {volume} {383}},\ \bibinfo {pages} {417} (\bibinfo {year}
  {2008})}\BibitemShut {NoStop}%
\bibitem [{\citenamefont {{Bekenstein}}(2004)}]{TeVeS}%
  \BibitemOpen
  \bibfield  {author} {\bibinfo {author} {\bibfnamefont {J.~D.}\ \bibnamefont
  {{Bekenstein}}},\ }\href@noop {} {\bibfield  {journal} {\bibinfo  {journal}
  {Phys. Rev. D}\ }\textbf {\bibinfo {volume} {70}},\ \bibinfo {pages} {083509}
  (\bibinfo {year} {2004})}\BibitemShut {NoStop}%
\bibitem [{\citenamefont {{Brownstein}}\ and\ \citenamefont
  {{Moffat}}(2006)}]{brownsteinmoffat}%
  \BibitemOpen
  \bibfield  {author} {\bibinfo {author} {\bibfnamefont {J.~R.}\ \bibnamefont
  {{Brownstein}}}\ and\ \bibinfo {author} {\bibfnamefont {J.~W.}\ \bibnamefont
  {{Moffat}}},\ }\href {\doibase 10.1086/498208} {\bibfield  {journal}
  {\bibinfo  {journal} {Astrophys. J.}\ }\textbf {\bibinfo {volume} {636}},\
  \bibinfo {pages} {721} (\bibinfo {year} {2006})}\BibitemShut {NoStop}%
\bibitem [{\citenamefont {{Milgrom}}(2009{\natexlab{b}})}]{bimond}%
  \BibitemOpen
  \bibfield  {author} {\bibinfo {author} {\bibfnamefont {M.}~\bibnamefont
  {{Milgrom}}},\ }\href {\doibase 10.1103/PhysRevD.80.123536} {\bibfield
  {journal} {\bibinfo  {journal} {Phys. Rev. D}\ }\textbf {\bibinfo {volume}
  {80}},\ \bibinfo {pages} {123536} (\bibinfo {year}
  {2009}{\natexlab{b}})}\BibitemShut {NoStop}%
\bibitem [{\citenamefont {{Mannheim}}\ and\ \citenamefont
  {{O'Brien}}(2010)}]{mannheimobrien}%
  \BibitemOpen
  \bibfield  {author} {\bibinfo {author} {\bibfnamefont {P.~D.}\ \bibnamefont
  {{Mannheim}}}\ and\ \bibinfo {author} {\bibfnamefont {J.~G.}\ \bibnamefont
  {{O'Brien}}},\ }\href@noop {} {\bibfield  {journal} {\bibinfo  {journal}
  {Phys. Rev. Lett.}\ ,\ \bibinfo {pages} {submitted}} (\bibinfo {year}
  {2010})},\ \Eprint {http://arxiv.org/abs/1007.0970} {arXiv:1007.0970}
  \BibitemShut {NoStop}%
\bibitem [{\citenamefont {{Blanchet}}(2007)}]{blanchet}%
  \BibitemOpen
  \bibfield  {author} {\bibinfo {author} {\bibfnamefont {L.}~\bibnamefont
  {{Blanchet}}},\ }\href {\doibase 10.1088/0264-9381/24/14/001} {\bibfield
  {journal} {\bibinfo  {journal} {Classical and Quantum Gravity}\ }\textbf
  {\bibinfo {volume} {24}},\ \bibinfo {pages} {3529} (\bibinfo {year}
  {2007})}\BibitemShut {NoStop}%
\bibitem [{\citenamefont {{Bruneton}}\ \emph {et~al.}(2009)\citenamefont
  {{Bruneton}}, \citenamefont {{Liberati}}, \citenamefont {{Sindoni}},\ and\
  \citenamefont {{Famaey}}}]{hybridDM}%
  \BibitemOpen
  \bibfield  {author} {\bibinfo {author} {\bibfnamefont {J.}~\bibnamefont
  {{Bruneton}}}, \bibinfo {author} {\bibfnamefont {S.}~\bibnamefont
  {{Liberati}}}, \bibinfo {author} {\bibfnamefont {L.}~\bibnamefont
  {{Sindoni}}}, \ and\ \bibinfo {author} {\bibfnamefont {B.}~\bibnamefont
  {{Famaey}}},\ }\href {\doibase 10.1088/1475-7516/2009/03/021} {\bibfield
  {journal} {\bibinfo  {journal} {J. Cosm. Astroparticle Phys.}\ }\textbf
  {\bibinfo {volume} {3}},\ \bibinfo {pages} {21} (\bibinfo {year}
  {2009})}\BibitemShut {NoStop}%
\bibitem [{\citenamefont {{Zhao}}\ and\ \citenamefont
  {{Li}}(2010)}]{darkfluid}%
  \BibitemOpen
  \bibfield  {author} {\bibinfo {author} {\bibfnamefont {H.}~\bibnamefont
  {{Zhao}}}\ and\ \bibinfo {author} {\bibfnamefont {B.}~\bibnamefont {{Li}}},\
  }\href {\doibase 10.1088/0004-637X/712/1/130} {\bibfield  {journal} {\bibinfo
   {journal} {Astrophys. J.}\ }\textbf {\bibinfo {volume} {712}},\ \bibinfo
  {pages} {130} (\bibinfo {year} {2010})}\BibitemShut {NoStop}%
\end{thebibliography}%

\end{document}